\tolerance=10000
\raggedbottom

\baselineskip=15pt
\parskip=1\jot

\def\sk{\vskip 3\jot}

\def\heading#1{\vskip3\jot{\noindent\bf #1}}
\def\label#1{{\noindent\it #1}}
\def\QED{\hbox{\rlap{$\sqcap$}$\sqcup$}}


\def\ref#1;#2;#3;#4;#5.{\item{[#1]} #2,#3,{\it #4},#5.}
\def\refinbook#1;#2;#3;#4;#5;#6.{\item{[#1]} #2, #3, #4, {\it #5},#6.} 
\def\refbook#1;#2;#3;#4.{\item{[#1]} #2,{\it #3},#4.}


\def\abs#1{\vert#1\vert}

\def\({\bigl(}
\def\){\bigr)}

\def\de{\delta}
\def\ep{\varepsilon}

\def\De{\Delta}

\def\yes{{\tt yes}}
\def\no{{\tt no}}

\def\Bool{\{0,1\}}
\def\Nset{\{1,\ldots, n\}}

\def\calD{{\cal D}}
\def\calF{{\cal F}}
\def\calG{{\cal G}}

\def\Booln{{\cal B}} 
\def\Dep{{\cal J}} 
\def\Sym{{\cal S}} 
\def\Const{{\cal C}} 
\def\QSym{{\cal Q}} 

{
\pageno=0
\nopagenumbers
\rightline{\tt testing.07.rev.tex}
\vskip0.5in

\centerline{\bf Attribute Estimation and Testing Quasi-Symmetry}
\vskip0.25in

\centerline{Krzysztof Majewski}
\centerline{\tt krzys@ifi.uio.no}
\centerline{Department of Informatics}
\centerline{University of Oslo}
\centerline{Boks 1072 Blindern}
\centerline{NO-0316 Oslo, Norway}
\vskip0.25in

\centerline{Nicholas Pippenger}
\centerline{\tt njp@hmc.edu}
\centerline{Department of Mathematics}
\centerline{Harvey Mudd College}
\centerline{1250 North Dartmouth Avenue}
\centerline{Claremont, CA 91711 USA}
\vskip0.25in

\noindent{\bf Abstract:}
A Boolean function is {\it symmetric\/} if it is invariant under all permutations of its arguments; it is 
{\it quasi-symmetric\/} if it is symmetric with respect to the arguments on which it actually depends.
We present a test that accepts every quasi-symmetric function and, except with an error probability at most $\de>0$,
rejects every function that differs from
every quasi-symmetric function on at least a fraction $\ep>0$ of the inputs.
For a function of $n$ arguments,
the test probes the function at 
$O\((n/\ep)\log(n/\de)\)$ 
inputs.
Our quasi-symmetry test acquires information concerning the arguments on which the function actually depends.
To do this, it employs a generalization of the property testing paradigm that we call
{\it attribute estimation}.
Like property testing, attribute estimation uses random sampling to obtain results that have
only ``one-sided'' errors and that are close to accurate with high probability.
\vfill\eject
}

\heading{1. Introduction}

Suppose that we are given a Boolean function $f:\Bool^n \to \Bool $ of $n$ arguments, and that we wish to determine whether or not it is symmetric. (Such a function $f$ is {\it symmetric} if it is invariant under all $n!$ permutations of its arguments: $f(x_1, \ldots, x_n) = f(x_{\pi(1)},\ldots, x_{\pi(n)})$ for all 
$(x_1, \ldots, x_n)\in \Bool^n$ and all bijections $\pi : \Nset \to \Nset$.
Equivalently, $f$ is symmetric if it is constant on sets of settings of its arguments
that have equal Hamming weight: $f(x) = f(y)$ if $\abs{x} = \abs{y}$, where the {\it Hamming weight\/}
$\abs{x}$ of $x\in\Bool^n$ is the number $\sum_{1\le i\le n} x_i$ of $1$s among $x_1, \ldots, x_n$.)
We seek to do this by querying the value of the function at various points (settings of its argument values), and we endeavor to minimize the number of queries that we make.

It is not hard to see that in the worst case, we may need to make $2^n - 2$ queries.
For if $f$ is the ``all $0$s'' constant function, we are bound to find that it is symmetric, but we dare not
announce this conclusion before querying all $2^n - 2$ points other than ``all $0$s''
$(0,\ldots,0)\in\Bool^n$ and ``all $1$s'' $(1,\ldots,1)\in\Bool^n$.
(For if $(x_1,\ldots, x_n)$ is such a point not queried, then the function $g$ that assumes the value $1$
at and only at the point $(x_1,\ldots, x_n)$ would be a non-symmetric function that assumes the same value as $f$ at all queried points.)

This situation may be summarized by saying that a ``witness'' for symmetry (a set of points at which the values of a function ensure its symmetry) must contain $2^n - 2$ points.
A witness for non-symmetry, however, may consist of just two points, $x$ and $y$, such that 
$\abs{x} = \abs{y}$, but for which $f(x) \not= f(y)$.
This suggests that we may test for symmetry more efficiently if we are willing to allow a small
probability of a ``false positive'' (when we declare a function to be symmetric when in fact it is not).

This suggestion leads us to the paradigm of  ``property testing''.
To introduce this notion to the current context, we shall need some definitions.
If $f$ and $g$ are Boolean functions of $n$ arguments, the {\it distance\/} $\De(f,g)$ between them
is the fraction of their truth-table entries on which they differ.
This fraction may be written as $\De(f,g) = \abs{f \oplus g}/2^n$, where $f\oplus g$ denotes the 
``exclusive-or'' or ``sum modulo $2$'' of $f$ and $g$, and the {\it Hamming weight\/} $\abs{f}$
of a Boolean function $f$ is the number $\sum_{x\in\Bool^n} f(x)$ of $1$s in its truth-table.
This is a proper metric on the set $\Booln_n$ of Boolean functions of $n$ arguments (that is, it is non-negative, vanishes only when the functions are equal, is symmetric and satisfies the triangle inequality).
Furthermore, if we extend it by agreeing that functions of different numbers of arguments
are at ``infinite distance'', it remains a metric, now defined on the set 
$\Booln = \bigcup_{n\ge 0} \Booln_n$ of all Boolean functions.
If $\calF$ and $\calG$ are sets of Boolean functions, we define
$\De(f,\calG) = \min_{g\in \calG} \De(f,g)$ and $\De(\calF,\calG) = \min_{f\in \calF} \De(f,\calG)$.
If $\De(f,\calG)\ge \ep$, we shall say that $f$ is {\it $\ep$-far\/} from $\calG$.

By a {\it test\/} for some property of Boolean functions
(reified as a set $\calG$ of Boolean functions), we shall mean a randomized algorithm that
takes a Boolean function $f$ and two parameters $\ep>0$ and $\de>0$, queries the value of 
$f$ at a finite sequence of points (where the number of queries, and choice of later points may depend on the outcomes of earlier queries), and announces an answer {\yes} or {\no}, where
(1) if $f$ has the property in question (that is, $f\in \calG$), then the algorithm answers \yes{}, and (2) if $f$ is $\ep$-far from the set of functions with the property in question (that is, $\De(f,\calG)\ge \ep$), then the algorithm answers \no{} unless an event of probability at most $\de$ has occurs.
(Note that ``probability'' here refers to the randomization of the algorithm; the bound $\de$ applies
uniformly to all $f$ that are $\ep$-far from $\calG$.)

Property testing, in the sense used here, was introduced by Rubinfeld and Sudan [R] and
further developed by Goldreich, Goldwasser and Ron [G3].
The definition has many variants and has been applied to properties of many types of objects.
A central example involving Boolean functions is testing monotonicity
(see Goldreich, Goldwasser, Lehman and Ron (and, in the later version, Samorodnitsky) [G1, G2]).

In Section 2, we shall give a test for symmetry that makes at most
$O\((1/ \ep)\log(1/\de)\)$ queries.
Although the test is randomized, this bound on the number of queries is uniform (not merely a bound on 
expectation, nor merely one that holds with high probability), and it is independent of the number of arguments.
When the algorithm returns \no, it also provides a witness to the non-symmetry of $f$.
(Of course, when \yes{} is returned, it may be a false positive, and in any case a witness to symmetry would be too large establish within the stated number of queries.)

We shall say that a Boolean function $f:\Bool^n \to \Bool $ of $n$ arguments is {\it quasi-symmetric\/}
if it is a symmetric function of those of its arguments that it actually depends on.
(We say that $f$ {\it depends on\/} its $i$-th argument if there are Boolean values
$x_1, \ldots, x_{i-1}, x_{i+1}, \ldots, x_n$ such that 
$f(x_1, \ldots, x_{i-1}, 0, x_{i+1}, \ldots, x_n)\not=f(x_1, \ldots, x_{i-1}, 1, x_{i+1}, \ldots, x_n)$.)

In Section 4, we shall give a test for quasi-symmetry that makes at most
$O\((n/ \ep)\log(n/\de)\)$ queries.
In this case, the number of queries depends on the number of $n$ of arguments as well as on
$\ep$ and $\de$.
This happens because the test begins by attempting to determine the set 
$\Dep(f)\subseteq \Nset$ of arguments
on which the function $f$ actually depends, and this set could be as large as the full set of $n$ arguments.

Upon considering the subproblem of determining the set $\Dep(f)$ from the function $f$, we see that it presents many of the
same characteristics as testing for symmetry: a witness for the fact that $f$ depends on its $i$-th arguments can comprise just two points (as exemplified in the definition), but a witness for the fact that 
$f$ does {\it not\/} depend on some particular argument cannot be smaller than all $2^n$ points.
Thus, instead of determining $\Dep(f)$ exactly, we shall introduce a notion of  ``estimating'' such an attribute that is exactly analogous to the notion of ``testing'' a property.

A function $\calD: \Booln \to C$  defined on the set 
of all Boolean functions, and taking values in a partially ordered set $C$,
will be called an {\it attribute\/} of Boolean functions.
An {\it estimate\/} for the attribute $\calD$ is a randomized algorithm that takes a Boolean function $f$ and two parameters $\ep>0$ and $\de>0$, queries the value of 
$f$ at a finite sequence of points (where the number of queries, and choice of later points may depend on the outcomes of earlier queries), and announces an output $D\in C$, where
(1) $\calD(f)\ge D$, and (2)  $f$ is $\ep$-far from the set $\calD^{-1}(D)$ of functions for which $\calD$ takes on the value $D$ only if an event of probability at most $\de$ occurs.
If we take $C = \{\yes, \no\}$ with $\yes < \no$, then estimating such an attribute reduces to testing
the corresponding property $\calD^{-1}(\yes)$.

In Section 3 we shall give an estimate for the set $\Dep(f)$ of arguments that $f$ depends on
 (with the codomain, the power set of 
$\Nset$, ordered in the usual way by inclusion) that makes at most 
$O\((n/ \ep)\log(n/\de)\)$ queries.
The algorithm will also provide a witness for the fact that the value it returns is a lower bound for
$\Dep(f)$.
This estimate will be used in Section 4 as the basis for our quasi-symmetry test, and the witness that it provides will be used to construct a witness for non-quasi-symmetry when that is detected.

The problem of testing whether a Boolean function depend on a small subset of arguments
has been attacked by Parnas, Ron and Samordnitsky [P1, P2], and later by
Fischer, Kindler, Ron, Safra and Samorodnitsky [F].
Their work, however, lies entirely within the framework of property testing: their algorithms
test whether the cardinality $\#\Dep(f)$ is at most $k$ ($k$ a constant), without giving further information about the set $\Dep(f)$
(and they do this with a number of queries that is independent of $n$).
Our quasi-symmetry test, however, requires more information about $\Dep(f)$ than merely its cardinality,
and this requirement led us to our formulation of the notion of attribute estimation.

Despite its naturalness and its analogy to property testing, attribute estimation does not appear to have
been described in the previous literature.
We hope, however, that it will find other applications, and indeed that the notion of attribute estimation,
both as used here and as extended to other domains such as graphs,
will prove a fruitful contribution to the theory of statistical algorithms.
\sk

\heading{2. Testing Symmetry}

In this section, we shall present our symmetry test and analyze its performance.
We shall begin by describing what we shall call a ``basic step'' of the algorithm.

\label{Symmetry Test---Basic Step:}
(0) Given Boolean function $f:\Bool^n \to \Bool$ of $n$ arguments,
return \yes{} if $0\le n\le 1$.
(1) Choose point $x = (x_1,\ldots, x_n)$ at random, with all $2^n - 2$ points other than
$(0,\ldots,0)$ and $(1,\ldots,1)$ being equally likely.
(2) Choose point $y = (y_1,\ldots, y_n)$ at random, with all ${n\choose \abs{x}} - 1$ points
other than $x$, but having the same Hamming weight as $x$, being equally likely.
(Since $n\ge 2$ and $x\not\in\{(0,\ldots,0), (1,\ldots, 1)\}$, there is at least one possible choice
for $y$.)
(3) Query $f(x)$ and $f(y)$.
If $f(x)=f(y)$, then return \yes, else return \no.

Clearly, if $f$ is symmetric, then this procedure returns \yes.
Let $\Sym$ denote the set of all symmetric Boolean functions.
We shall see that if $f$ is not symmetric, then the procedure returns \no{} with probability at least
$\De(f,\Sym)$.
To see this, let $A\subseteq\Bool^n$ be a set of points of minimum 
cardinality  such that
complementing the value of $f$ at just those points in $A$ yields a symmetric function.
This minimum cardinality is $\#A = \De(f,\Sym)\,2^n$.
For any $x\in \Bool^n$, let $B_x\subseteq\Bool^n$ denote the set of points $y$ such that 
$\abs{y}=\abs{x}$ but $f(y)\not=f(x)$.
The probability that the basic step chooses $x\in A$ is $\De(f,\Sym)\,2^n/(2^n-2) \ge \De(f,\Sym)$.
Given that $x\in A$, the probability that the basic step chooses $y\in B_x$ is
at least ${1\over 2}{n\choose \abs{x}}/\left({n\choose \abs{x}}-1\right) \ge {1\over 2}$
(since if  $\#B_x$ were less than ${1\over 2}{n\choose \abs{x}}$, we could replace
the points of $A$ having Hamming weight $\abs{x}$ by those of $B_x$, reducing the cardinality of $A$
and contradicting the definition of $A$).
The procedure returns \no{} if $x\in A$ and $y\in B_x$, which occurs with probability at least
${1\over 2}\De(f,\Sym)$, and also if $y\in A$ and $x\in B_y$, which occurs disjointly with a probability 
that is also at least
${1\over 2}\De(f,\Sym)$ (since the joint distribution of $x$ and $y$ is invariant under the exchange of 
$x$ and $y$).
This completes the proof that the basic step returns \no{} with probability at least $\De(f,\Sym)$.

Now we present our complete symmetry test.

\label{Symmetry Test:}
(0) Given Boolean function $f:\Bool^n \to \Bool$ of $n$ arguments, and real numbers
$0<\ep<1$ and $0<\de<1$, if $0\le n\le 1$, then return \yes.
(1)  Otherwise compute
$$k = \left\lceil {1\over \ep}\log{1\over \de}\right\rceil.$$
(2) Perform the basic step until it returns \no, or until it has returned \yes{} $k$ times.
(3) If any performance of the basic step returned \no, then return \no; if all $k$ performances
returned \yes, then return \yes.

Clearly, if $f$ is symmetric, then this procedure returns \yes.
We shall see that if $f$ is $\ep$-far from the symmetric functions, then the procedure returns \no{} with probability at least $1 - \de$.
If $f$ is $\ep$-far from the symmetric functions, then each performance of the basic step returns \yes{} with probability at most $1-\De(f,\Sym) \le 1 - \ep$.
The symmetry test returns \yes{} only if $k$ performances of the basic step return \yes, and this occurs with probability at most $(1 - \ep)^k \le e^{-\ep k} \le \de$ (where we have used the inequality 
$1 + x \le e^x$ and the definition of $k$).
This completes the proof that, if  if $f$ is $\ep$-far from the symmetric functions, then the procedure returns \no{} with probability at least $1 - \de$.

We now have the following theorem.

\label{Theorem 2.1:}
There is a test for symmetry that makes at most
$O\((1/ \ep)\log(1/\de)\)$ queries.

\label{Proof:}
The number of queries made is at most
$$2k = O\left( {1\over\ep} \log {1\over \de} \right).$$
\QED

We conclude this section by observing that when the basic step returns \no, the two points it has queried
witness the non-symmetry of the function, and when the complete symmetry test returns \no,
such a  witness is provided by the final performance of the basic step.
\sk

\heading{3. Estimating Dependence}

In this section we shall present an estimate,
as defined in the introduction, for the set $\Dep(f)$ of arguments that $f$ depends on.
We begin by describing a constancy test analogous to the symmetry test presented in the preceding section.

\label{Constancy  Test---Basic Step:}
(0) Given Boolean function $f:\Bool^n \to \Bool$ of $n$ arguments,
return \yes{} if $n=0$.
(1) Choose point $x = (x_1,\ldots, x_n)$ at random, with all $2^n$ points  being equally likely.
(2) Choose point $y = (y_1,\ldots, y_n)$ at random, with all $2^n - 1$ points
other than $x$ being equally likely.
(3) Query $f(x)$ and $f(y)$.
If $f(x)=f(y)$, then return \yes, else return \no.

Clearly, if $f$ is constant, then this procedure returns \yes.
Let $\Const$ denote the set of all constant Boolean functions.
We shall see that if $f$ is not constant, then the procedure returns \no{} with probability at least
$\De(f,\Const)$.
To see this, let $A\subseteq\Bool^n$ be a set of points of minimum cardinality  such that
complementing the value of $f$ at just those points in $A$ yields a constant function.
This minimum cardinality is $\#A = \De(f, \Const)\,2^n$.
For any $x\in \Bool^n$, let $B_x\subseteq\Bool^n$ denote the set of points $y$ such that 
$f(y)\not=f(x)$.
The probability that the basic step chooses $x\in A$ is $\De(f, \Const)$.
Given that $x\in A$, the probability that the basic step chooses $y\in B_x$ is
at least $2^{n-1}/(2^n - 1) \ge {1\over 2}$
(since if  $\#B_x$ were less than $2^{n-1}$, we could replace
the points of $A$ by those of $B_x$, reducing the cardinality of $A$
and contradicting the definition of $A$).
The procedure returns \no{} if $x\in A$ and $y\in B_x$, which occurs with probability at least
${1\over 2}\De(f, \Const)$, and also if $y\in A$ and $x\in B_y$, which occurs disjointly with a probability that is at least
${1\over 2}\De(f, \Const)$ (since the joint distribution of $x$ and $y$ is invariant under the exchange of 
$x$ and $y$).
This completes the proof that the basic step returns \no{} with probability at least $\De(f, \Const)$.

Now we present our complete constancy test.

\label{Constancy Test:}
(0) Given Boolean function $f:\Bool^n \to \Bool$ of $n$ arguments, and real numbers
$0<\ep<1$ and $0<\de<1$, if $n=0$, then return \yes.
(1) Otherwise compute
$$k = \left\lceil {1\over \ep}\log{1\over \de}\right\rceil.$$
(2) Perform the basic step until it returns \no, or until it has returned \yes{} $k$ times.
(3) If any performance of the basic step returned \no, then return \no; if all $k$ performances
returned \yes, then return \yes.

Clearly, if $f$ is constant, then this procedure returns \yes.
We shall see that if $f$ is $\ep$-far from the constant functions, then the procedure returns \no{} with probability at least $1 - \de$.
If $f$ is $\ep$-far from the constant functions, then each performance of the basic step returns \yes{} with probability at most $1-\De(f, \Const) \le 1 - \ep$.
The complete test returns \yes{} only if $k$ performances of the basic step return \yes, and this occurs with probability at most $(1 - \ep)^k \le e^{-\ep k} \le \de$ (where we have used the inequality 
$1 + x \le e^x$ and the definition of $k$).
This completes the proof that, if  if $f$ is $\ep$-far from the constant functions, then the procedure returns \no{} with probability at least $1 - \de$.

We now have the following lemma.

\label{Lemma 3.1:}
There is a test for constancy that makes at most
$O\((1/ \ep)\log(1/\de)\)$ queries.

\label{Proof:}
The number of queries made is at most
$$2k = O\left( {1\over\ep} \log {1\over \de} \right).$$
\QED

We observe that when the basic step returns \no, the two points it has queried
witness the non-constancy of the function, and when the complete constancy test returns \no,
such a  witness is provided by the final performance of the basic step.

The next component we shall need is a procedure that takes a pair of points that witness the non-constancy of a function and returns a particular argument on which the function depends.
We shall call this operation a ``dependency search''.

\label{Dependency Search:}
(0) Given Boolean function $f:\Bool^n \to \Bool$ of $n$ arguments, and a pair of points $x, y\in\Bool^n$
such that $f(x)\not=f(y)$, set $x' := x$ and $y' := y$.
(1) If $\abs{x'\oplus y'} = 1$, then return the unique $i\in\{1,\ldots,n\}$ such that
$x'_i \not= y'_i$.
(2) Otherwise, let $z\in\Bool^n$ be such that $\abs{x'\oplus z}$ and $\abs{y'\oplus z}$ are each at most
$\lceil \abs{x'\oplus y'} / 2 \rceil$ (so that $z$ is about half-way between $x'$ and $y'$).
(3)  Query $f(z)$.
If $f(x')\not=f(z)$, then set $y' := z$, else set $x' := z$.
(4) Go back to step (1).

\label{Lemma 3.2:}
There is a procedure for dependency search that makes $O(\log n)$ queries.

\label{Proof:}
In the procedure given above, the quantity $\abs{x'\oplus y'}$ is initially at most $n$.
The procedure  reduces this quantity by multiplying it by
a factor at most $2/3$ whenever it makes a query, and stops when this quantity reaches $1$. 
Thus it makes at most $\log_{3/2} n$ queries.
\QED

We observe that the final values of $x'$ and $y$ in this dependency search provide a witness that
$f$ actually depends on the $i$-th argument.

We now present our complete dependency estimate.

\label{Dependency Estimate:}
(0) Given a Boolean function $f : \Bool^n \to \Bool$ of $n$ arguments, and real numbers
$0<\ep<1$ and $0<\de<1$, let $\de' = \de/n$  and set $J' := \emptyset$.
(1) Set $x_{J'} = \{x_j : j\in J'\}$ to random Boolean values,
with all $2^{\#J'}$ assignments being equally likely, and let $f' : \Bool^{n-\#J'} \to \Bool$ be the 
Boolean function obtained from $f$ by substituting the values $x_{J'}$ for the arguments 
in $J'$.
(2) Perform a constancy test on the function $f'$ with parameters $\ep$ and $\de'$.
(3) If this test returns \yes, then return $J'$.
(4) Otherwise, perform a dependency search on the witness returned by the constancy test,
and adjoin the argument $j$ returned by the dependency search to $J'$:
$J' := J' \cup\{j\}$.
(5) Go back to step (1). 

If $J$ denotes the set returned by this procedure, it is clear that $J \subseteq \Dep(f)$.
Indeed, this fact is witnessed by the set of pairs of points  $x', y'$ that terminate the dependency searches that found the arguments in $J$.
Let $\Booln_J$ be the set of Boolean functions that actually depend only on the arguments in $J$.
We shall show that that if $f$ is $\ep$-far from any function in $\Booln_J$, then an event of probability
at most $\de$ has occurred.
If $f$ is $\ep$-far from any function in $\Booln_J$, we have $\De\(f, \Booln_J\) \ge \ep$.
For each performance of the constancy test, we have $J' \subseteq J$; this implies
$\De\(f, \Booln_{J'}\) \ge\De\(f, \Booln_J\)$, and thus we have
$\De\(f, \Booln_J\) \ge \ep$.
One of these at most $n$ constancy tests must return \yes, and each test returns \yes{} with probability
at most $\de' = \de/n$.
Thus the probability that any of these tests returns \yes{} is at most $n\de' = \de$.
Finally, the number of queries made is at most
$$n\,\left( O\left( {1\over \ep} \log {1\over \de'}\right) + O(\log n) \right) =
n\,\left( O\left( {1\over \ep} \log {n\over \de}\right) + O(\log n) \right) =
O\left( {n\over \ep} \log {n\over \de}\right).$$

We now have the following theorem.

\label{Theorem 3.3:}
There is an estimate for the dependency set that makes at most 
$O\((n/ \ep)\log(n/\de)\)$ queries.
\sk

\heading{4. Testing Quasi-Symmetry}

In this section, we present our quasi-symmetry test.
Again we begin by describing a basic step.

\label{Quasi-Symmetry Test---Basic Step:}
(0) Given a Boolean function $f : \Bool^n \to \Bool$ of $n$ arguments, a real number $0<\ep<1$, and a set $J$ of arguments, set $I = \{1,\ldots, n\}\setminus J$.
(1) Set $x_I = \{x_i : i\in I\}$ to random Boolean values,
with all $2^{\#I}$ assignments being equally likely, and let $f' : \Bool^{\#J} \to \Bool$ be the 
Boolean function obtained from $f$ by substituting the values $x_I$ for the arguments 
in $I$.
(2) Perform a symmetry test on $f'$, with parameters $\ep$ and $1/2$, and return the value
returned by this symmetry test as the value of the basic step.

Suppose that the basic step is performed on a function $f$ that  actually
 depends on all the arguments in $J$.
 If $f$ is quasi-symmetric, then because $f$ actually depends on all the arguments in $J$
and $f'$ depends only on these arguments, $f'$ is symmetric and the basic step will return \yes.
Let $g\in\Booln_J$ be a function that minimizes $\De(f,g)$.
This minimum distance is $\De(f,g) = \De(f,\Booln_J)$.
Now suppose further that
$\De(f,g)\le \ep$.
Then we shall show that
if $f$ is $4\ep$-far from the set $\QSym$ of quasi-symmetric  functions,  the basic step returns \no{} with probability at least $1/4$.
By the triangle inequality we have $\De(f,g) + \De(g,\QSym) \ge \De(f,\QSym) \ge 4\ep$,
and thus $\De(g, \QSym) \ge 3\ep$.
Let $g'$ be obtained from $g$ by substituting the random values $x_I$ for the arguments in $I$.
Of course $g'$ is independent of these random values, since $g$ does not depend on the arguments in $I$.
Furthermore, $\De(g',\Sym) \ge \De(g',\QSym) \ge \De(g,\QSym) \ge 3\ep$, since $\Sym\subseteq\QSym$ and $g$ does not depend on any of its arguments that are not arguments of $g'$.  
On the other hand, $\De(g',f')$ is a random variable, since $f'$ may depend on the random values $x_I$.
The expected value of $\De(g',f')$ is clearly $\De(g,f) = \De(f,g) \le \ep$, so with probability at least $1/2$ we have $\De(f',g') \le 2\ep$, by Markov's inequality.
When this happens, we have $\De(g',f') + \De(f',\Sym) \ge \De(g',\Sym) \ge 3\ep$, again by the triangle
inequality, and thus $\De(f',\Sym)\ge \ep$.
This condition ensures that the symmetry test returns \no{} with probability at least $1/2$.
This completes the proof that when $f$ is $4\ep$-far from the set $\QSym$ of quasi-symmetric  functions, then the basic step returns \no{} with probability at least $1/4$.

\label{Quasi-Symmetry Test:}
(0) Given a Boolean function $f : \Bool^n \to \Bool$ of $n$ arguments, and real numbers
$0<\ep<1$ and $0<\de<1$, let $\ep' = \ep / 4$ and $\de' = \de / 2$,
and compute
$$k = \left\lceil \log_{4/3} {1\over \de' }\right\rceil.$$
(1) Perform a dependency estimate on $f$, with parameters $\ep'$ and $\de'$, and let $J$ be 
the value returned by that procedure.
(2) Perform the basic step with parameters $f$, $\ep'$, and $J$,
until it returns \no, or until it has returned \yes{} $k$ times.
(3) If any performance of the basic step returned \no, then return \no; if all $k$ performances
returned \yes, then return \yes.

Suppose first that $f$ is quasi-symmetric.
The function $f$ actually depends on all the arguments in the set $J$ found in step (1).
Thus each performance of the basic step returns \yes, and so the quasi-symmetry test returns
\yes{} after $k$ such performances.
Suppose on the other hand that $f$ is $\ep$-far from any quasi-symmetric function.
Then, except with probability at most $\de'$, $\De\(f,\Booln_J\) \le \ep'$ for the set $J$ found by
step (1).
Thus each performance of the basic step returns $\no$ with probability at least $1/4$,
and therefore returns \yes{} with probability at  most $3/4$.
The probability that all $k$ performances of the basic step return \yes{} is thus at most $(3/4)^k \le \de'$.
Thus if $f$ is $\ep$-far from the quasi-symmetric functions, the quasi-symmetry test
returns \no{} unless an event of probability at most $\de' + \de' = \de$ occurs.

Finally, the number of queries made is at most
$$O\left( {n\over \ep'} \log {n\over \de'} \right) + O\left( {1\over \ep'} \log {1\over \de'} \right) = 
O\left( {4n\over \ep} \log {2n\over \de} \right) + O\left( {4\over \ep} \log {2\over \de} \right) =
O\left( {n\over \ep} \log {n\over \de} \right).$$

We now have the following theorem.

\label{Theorem 4.3:}
There is a test for quasi-symmetry that makes at most 
$O\((n/ \ep)\log(n/\de)\)$ queries.

We observe that when the quasi-symmetry test returns \no{}, a witness to the non-quasi-symmetry
of $f$ can be obtained from the witnesses provided by the dependency estimate and the final
performance of the symmetry test.
\sk

\heading{5. Acknowledgments}

Preliminary versions of the results presented here appeared in the first author's thesis, and were supported by a Canada Research Chair and a Discovery Grant to the second author from  the National Science and Engineering Research Council of Canada.
Subsequent work was supported by Grant CCF 043056 to the second author by the National Science Foundation of the United States.
\sk

\heading{6. References}

\ref F; E. Fischer, G. Kindler, D. Ron, S. Safra and A. Samoridnitsky;
``Testing Juntas'';
Proc.\ IEEE Symp.\ on Foundations of Computer Science; 43 (2002) 103--112.

\ref G1; O. Goldreich, S. Goldwasser, E. Lehman and D. Ron;
``Testing Monotonicity'';
Proc.\ IEEE Symp.\ on Foundations of Computer Science; 39 (1998) 426--435.

\ref G2; O. Goldreich, S. Goldwasser, E. Lehman, D. Ron and A. Samoridnitsky;
``Testing Monotonicity'';
Combinatorica; 20 (2000) 301--337.

\ref G3; O. Goldreich, S. Goldwasser and D. Ron;
``Property Testing and Its Connection to Learning and Approximation'';
J. ACM; 45 (1998) 653--750.

\refbook M; K. Majewski;
Probabilistic Testing of Boolean Functions; 
M.~Sc. Thesis, Department of Computer Science, University of British Columbia, 2003.

\refinbook P1; M. Parnas, D. Ron and A. Samoridnitsky;
``Proclaiming Dictators and Juntas or Testing Boolean Formulae'';
in: M.~Goemans, K.~Jansen, J.~D.~P. Rolim and L.~Treviasan (Ed's);
Approximation, Randomization, and Combinatorial Optimization;
{Lecture Notes in Computer Science, v.~2129, Springer-Verlag, Berlin, 2001, pp.~273--284}.

\ref P2; M. Parnas, D. Ron and A. Samoridnitsky;
``Proclaiming Dictators and Juntas or Testing Boolean Formulae'';
Electronic Colloq.\ on Computational Complexity; {{\tt http://eccc.uni-trier.de/eccc/}, 
2001, no.~63, 31~pp.}.

\ref P3; M. Parnas, D. Ron and A. Samoridnitsky;
``Testing Basic Boolean Formulae'';
Electronic Colloq.\ on Computational Complexity; {{\tt http://eccc.uni-trier.de/eccc/}, 
2001, no.~63, rev.~1, 29~pp.}.

\ref R; R. Rubinfeld and M. Sudan;
``Robust Characterizations of Polynomials with Applications to Program Testing'';
SIAM J. Computing; 25:2 (1996) 252-271.

\bye